\documentstyle[prb,aps,twocolumn,epsf]{revtex}

\newcommand{\beq}{\begin{equation}}
\newcommand{\eeq}{\end{equation}}
\newcommand{\bea}{\begin{eqnarray}}
\newcommand{\eea}{\end{eqnarray}}

\newcommand{\etal}{{\em et al.}}

\def\jour#1#2#3#4{{#1} {\bf #2}, #3 (#4)}
\def\tit#1#2#3#4#5{{#1} {\bf #2}, #3 (#4)}

\def\prl{Phys.\ Rev.\ Lett.\ }

\def\prb{Phys.\ Rev.\ B}

\def\cjp{Can.\ J.\ Phys.\ }
\def\jpsj{J.\ Phys.\ Soc.\ Jap.\  }
\def\jsp{J.\ Stat.\ Phys.\ }
\def\epjb{Eur.\ Phys.\ J.\ B}

\def\mr{m_{rms}}

\begin{document}
\draft

\twocolumn[\hsize\textwidth\columnwidth\hsize\csname @twocolumnfalse\endcsname

\title{The phase diagram of the hexagonal lattice quantum dimer model}

\author{R. Moessner,$^1$ S. L. Sondhi$^1$\ and P. Chandra$^2$}
\address{$^1$Department of Physics, Jadwin Hall, Princeton University,
Princeton, NJ 08544, USA\\ 
$^2$NEC Research Institute, 4 Independence
Way, Princeton NJ 08540, USA} 
\date{\today}

\maketitle

\begin{abstract}

We discuss the phase diagram of the quantum dimer model on the hexagonal 
(honeycomb) lattice. In addition to the columnar and staggered valence 
bond solids which have been discussed in previous work, we establish the 
existence of a plaquette valence bond solid. The transition between the
plaquette and columnar phases at $v/t=-0.2\pm0.05$\ is argued to be first 
order. We note that this model should describe valence bond dominated
phases of frustrated Heisenberg models on the hexagonal lattice and discuss
its relation to recent exact diagonalisation work by J.B. Fouet \etal\
on the $J_1-J_2$\ model on the same lattice.
Our results also shed light on the properties of the transverse 
field Ising antiferromagnet on the triangular lattice and the classical
Ising
antiferromagnet on the stacked triangular lattice, which are related
to dimer models by duality.
\end{abstract}

]

\section{Introduction}

Quantum dimer models (QDMs) were introduced to describe the physics of
Heisenberg antiferromagnets in a regime dominated by valence
bonds.\cite{Rokhsar88} This regime is best realised in cases where the
conventional Neel state is destabilised by quantum fluctuations or prohibited
by frustration. The most prominent appearance of such dimer models has
been in the context of the superconductivity of the cuprates, where
the QDM on the square lattice was introduced by Rokhsar and 
Kivelson\cite{Rokhsar88} to describe the physics of the short-range 
flavour\cite{kivrokset} of resonating valence bond physics.\cite{pwa87}
Subsequently, Read and Sachdev showed that QDMs arise naturally in
certain ``extreme quantum limits'' of generalizations of $SU(2)$
magnets to $SU(N)$ or $Sp(N)$ symmetry with large $N$.\cite{readsach2}

More recently, an exact duality between QDMs in $d=2$ and frustrated 
Ising models in a weak transverse field was explored by the
present authors.\cite{mcs2000} This
mapping connects the QDM not only to transverse field Ising models
but also to a class of ferromagnetically stacked frustrated Ising
magnets in dimension $d=2+1$, which are of independent interest
as they also have a range of experimental 
realisations.\cite{collinsrev}
The connection between QDMs, short-ranged RVB physics and Ising
gauge theories has also been discussed recently by Fradkin and
two of the present authors (RM and SLS).\cite{msf}

From the work to date, it appears that QDM on the square lattice
does not realise the disordered phase envisaged in the short ranged RVB 
scenario of high-temperature superconductivity. Rather, 
it is ordered everywhere except at a point, a possibility already noted in 
RK's original work,\cite{Rokhsar88} and fleshed out from various
viewpoints by different 
authors.\cite{subirfinitesize,ioffelar89,levitov,readsach2,leung,henleyheight,mcs2000}
While some evidence has been presented dissenting
from this scenario,\cite{orland,leung} as 
will become clear in the following, we believe there no longer is any
real
basis for doubting it.

The physics of the square lattice QDM is closely connected to the physics
of the hexagonal lattice QDM -- both lattices exhibit critical classical
dimer correlations which can be traced, via height representations, to
their bipartite nature. (In contrast the non-bipartite triangular lattice
exhibits disordered classical correlations and an RVB phase in its
QDM.\cite{mstridimer}) This connection was discussed first by Read
and Sachdev (see Ref.~\onlinecite{readsach2}) 
and has more recently been discussed 
by us in the context of a study of frustrated transverse field Ising 
models.\cite{mcs2000}

Whereas the general structure of the phase diagram for both
problems has been in place for some time, settling the detailed structure 
has turned out to be difficult. Approaches based on 
mappings to height models, or Landau-Ginzburg-Wilson theory contain 
undetermined
parameters upon which the detailed nature of the ordering pattern
depends. From a point of view of numerics, the problems of simulating
the quantum problem has been restrictive in that the studies thus far
have been limited to diagonalisations of systems of rather small
size,\cite{subirfinitesize,leung} from which subtle difference in
correlations have been difficult to read off.

In this paper, we solve this problem for the the QDM on the hexagonal 
(honeycomb) lattice and map out its phase diagram. This is done with
the aid of the abovementioned mapping to a {\em classical} stacked
triangular Ising magnet. This mapping gives access to a number of
analytical results but, more importantly, allows efficient numerical
simulations of systems much larger than the ones previously studied.

Our central result is that the QDM on the hexagonal lattice has
three phases, namely a staggered, a plaquette and a columnar valence
bond solid (VBS). The transitions between these phases are a
first order transition between the columnar and plaquette phases
and a combination of a first order and continuous transition in the 
case of the plaquette and staggered phases; the latter involves
fluctations in the ground state on one side of the transition
but not on the other.

We further discuss the implications of these results for the
properties of stacked triangular Ising magnets with nearest-neighbour
interactions,\cite{Blankschtein84,Netz84,Coppersmith85,mats87,heino89,kim90,Bunker95,plumer95,collinsrev}
for which our algorithm allows us to avoid some numerical limitations
encountered in previous studies, and where the extension to the QDM
provides considerable insight into the stability of the low temperature
phase. This phase turns out to be one with three inequivalent
sublattices, one of which is disordered, in accordance with
the results of some, but not all, previous studies.

We also review the connection of the QDM under study to frustrated
Heisenberg antiferromagnets on the hexagonal lattice. Such magnets
are prime candidate for being described by the quantum dimer model, 
and it turns out that the Heisenberg  model with competing interactions 
does indeed seem to realise the order present in two of the phases
of the QDM.\cite{fouet}

Turning to the QDM, its Hilbert space consists of hardcore dimer coverings of
the hexagonal lattice. The Hamiltonian acts on each hexagonal
plaquette of the lattice. It contains two terms, a kinetic ($\hat{T}$)
and a potential ($\hat{V}$) one. The former generates a plaquette
resonance move by rotating a triplet of dimers by 60$^\circ$ (see
Fig.~\ref{fig:phases}), in analogy to the benzene
resonance.\cite{benzene} The latter is diagonal in the dimer basis and
simply counts the number of plaquettes able to resonate (`flippable
plaquettes').

The Hamiltonian of the QDM can thus be represented as a
sum over plaquettes of the following plaquette Hamiltonian:
\bea H_{ QDM} &=&-t\hat{T}+v\hat{V}
\nonumber\\
&=& -t\left( | \setlength{\unitlength}{3158sp}%
\begingroup\makeatletter\ifx\SetFigFont\undefined%
\gdef\SetFigFont#1#2#3#4#5{%
  \reset@font\fontsize{#1}{#2pt}%
  \fontfamily{#3}\fontseries{#4}\fontshape{#5}%
  \selectfont}%
\fi\endgroup%
\begin{picture}(194,185)(318,260)
\thicklines \put(468,421){\circle{18}} 
\put(358,420){\line( 1, 0){108}} 
\put(342,377){\circle{18}} \put(360,421){\circle{18}}
\put(396,284){\circle{18}} 
\multiput(434,282)(5.48824,9.14707){11}{\makebox(8.3333,12.5000){\SetFigFont{7}{8.4}{\rmdefault}{\mddefault}{\updefault}.}}
\multiput(340,378)(5.57647,-9.29412){11}{\makebox(8.3333,12.5000){\SetFigFont{7}{8.4}{\rmdefault}{\mddefault}{\updefault}.}}
\put(434,284){\circle{18}}
\put(488,377){\circle{18}}
\end{picture}
 \rangle
\langle
\setlength{\unitlength}{3158sp}%
\begingroup\makeatletter\ifx\SetFigFont\undefined%
\gdef\SetFigFont#1#2#3#4#5{%
  \reset@font\fontsize{#1}{#2pt}%
  \fontfamily{#3}\fontseries{#4}\fontshape{#5}%
  \selectfont}%
\fi\endgroup%
\begin{picture}(194,185)(318,93)
\thicklines \put(468,117){\circle{18}} 
\put(358,118){\line( 1, 0){108}} 
\put(342,161){\circle{18}} \put(360,117){\circle{18}}
\put(396,254){\circle{18}} 
\multiput(434,256)(5.48824,-9.14707){11}{\makebox(8.3333,12.5000){\SetFigFont{7}{8.4}{\rmdefault}{\mddefault}{\updefault}.}}
\multiput(340,160)(5.57647,9.29412){11}{\makebox(8.3333,12.5000){\SetFigFont{7}{8.4}{\rmdefault}{\mddefault}{\updefault}.}}
\put(434,254){\circle{18}}
\put(488,161){\circle{18}}
\end{picture}
|+h.c.  \right) +v\left
( |
\setlength{\unitlength}{3158sp}%
\begingroup\makeatletter\ifx\SetFigFont\undefined%
\gdef\SetFigFont#1#2#3#4#5{%
  \reset@font\fontsize{#1}{#2pt}%
  \fontfamily{#3}\fontseries{#4}\fontshape{#5}%
  \selectfont}%
\fi\endgroup%
\begin{picture}(194,185)(318,260)
\thicklines \put(468,421){\circle{18}} 
\put(358,420){\line( 1, 0){108}} 
\put(342,377){\circle{18}} \put(360,421){\circle{18}}
\put(396,284){\circle{18}} 
\multiput(434,282)(5.48824,9.14707){11}{\makebox(8.3333,12.5000){\SetFigFont{7}{8.4}{\rmdefault}{\mddefault}{\updefault}.}}
\multiput(340,378)(5.57647,-9.29412){11}{\makebox(8.3333,12.5000){\SetFigFont{7}{8.4}{\rmdefault}{\mddefault}{\updefault}.}}
\put(434,284){\circle{18}}
\put(488,377){\circle{18}}
\end{picture}
\rangle \langle
\setlength{\unitlength}{3158sp}%
\begingroup\makeatletter\ifx\SetFigFont\undefined%
\gdef\SetFigFont#1#2#3#4#5{%
  \reset@font\fontsize{#1}{#2pt}%
  \fontfamily{#3}\fontseries{#4}\fontshape{#5}%
  \selectfont}%
\fi\endgroup%
\begin{picture}(194,185)(318,260)
\thicklines \put(468,421){\circle{18}} 
\put(358,420){\line( 1, 0){108}} 
\put(342,377){\circle{18}} \put(360,421){\circle{18}}
\put(396,284){\circle{18}} 
\multiput(434,282)(5.48824,9.14707){11}{\makebox(8.3333,12.5000){\SetFigFont{7}{8.4}{\rmdefault}{\mddefault}{\updefault}.}}
\multiput(340,378)(5.57647,-9.29412){11}{\makebox(8.3333,12.5000){\SetFigFont{7}{8.4}{\rmdefault}{\mddefault}{\updefault}.}}
\put(434,284){\circle{18}}
\put(488,377){\circle{18}}
\end{picture}
|+
|
\setlength{\unitlength}{3158sp}%
\begingroup\makeatletter\ifx\SetFigFont\undefined%
\gdef\SetFigFont#1#2#3#4#5{%
  \reset@font\fontsize{#1}{#2pt}%
  \fontfamily{#3}\fontseries{#4}\fontshape{#5}%
  \selectfont}%
\fi\endgroup%
\begin{picture}(194,185)(318,93)
\thicklines \put(468,117){\circle{18}} 
\put(358,118){\line( 1, 0){108}} 
\put(342,161){\circle{18}} \put(360,117){\circle{18}}
\put(396,254){\circle{18}} 
\multiput(434,256)(5.48824,-9.14707){11}{\makebox(8.3333,12.5000){\SetFigFont{7}{8.4}{\rmdefault}{\mddefault}{\updefault}.}}
\multiput(340,160)(5.57647,9.29412){11}{\makebox(8.3333,12.5000){\SetFigFont{7}{8.4}{\rmdefault}{\mddefault}{\updefault}.}}
\put(434,254){\circle{18}}
\put(488,161){\circle{18}}
\end{picture}
\rangle \langle
\setlength{\unitlength}{3158sp}%
\begingroup\makeatletter\ifx\SetFigFont\undefined%
\gdef\SetFigFont#1#2#3#4#5{%
  \reset@font\fontsize{#1}{#2pt}%
  \fontfamily{#3}\fontseries{#4}\fontshape{#5}%
  \selectfont}%
\fi\endgroup%
\begin{picture}(194,185)(318,93)
\thicklines \put(468,117){\circle{18}} 
\put(358,118){\line( 1, 0){108}} 
\put(342,161){\circle{18}} \put(360,117){\circle{18}}
\put(396,254){\circle{18}} 
\multiput(434,256)(5.48824,-9.14707){11}{\makebox(8.3333,12.5000){\SetFigFont{7}{8.4}{\rmdefault}{\mddefault}{\updefault}.}}
\multiput(340,160)(5.57647,9.29412){11}{\makebox(8.3333,12.5000){\SetFigFont{7}{8.4}{\rmdefault}{\mddefault}{\updefault}.}}
\put(434,254){\circle{18}}
\put(488,161){\circle{18}}
\end{picture}
|\right)\ .\nonumber 
\eea 
It has one free parameter, namely the ratio
of the strength of the potential and kinetic terms, $v/t$.

The structure of this paper is as follows. In Sect.~\ref{sect:cand},
we discuss the phases which one might expect to encounter in the model
under consideration. Sect.~\ref{sect:maps} contains a summary of the
methods used to establish the results that follow. The  
numerical results on the QDM are presented in Sect.~\ref{sect:numres},
from which the phase diagram (Sect.~\ref{sect:phasedia}) follows. 
We then discuss implications for the study of magnets, 
namely triangular stacked Ising (Sect.~\ref{sect:stacked}) and 
$S=1/2$\ hexagonal Heisenberg (Sect.~\ref{sect:hexag}) models.
We close with a conclusion in Sect.~\ref{sect:concl}.

\section{Candidate phases}
\label{sect:cand}

\begin{figure}
\epsfxsize=3in
\centerline{\epsffile{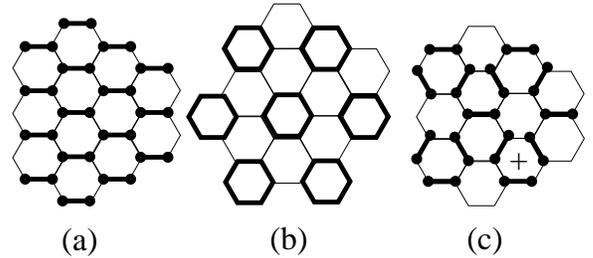}}
\caption{Dimer patterns on the hexagonal lattice: (a) staggered, (b)
plaquette and (c) columnar. The marked links have a high probability
of being occupied by a dimer in the respective phases.  Note that in
each case, there are only two inequivalent sets of links.  A dimer
plaquette move effected by $\hat{T}$\ consists of rotating the three
dimers surrounding a plaquette (like the one labelled with a plus) by
60$^\circ$. }
\label{fig:phases}
\end{figure}

As mentioned above, the QDM on the hexagonal lattice is closely
connected to its square lattice version. Hence a number of
known exact statements on the square lattice carry over
{\it mutatis mutandis} to the hexagonal lattice.
Firstly, for $v>t$, the ground state is
the staggered state, $\left|\varphi\right>$, depicted in
Fig.~\ref{fig:phases}a. This follows from the fact that a lower bound
on the energy per plaquette is $\min\{0,v-t\}$, and only
$\left|\varphi\right>$\ saturates this bound for $v>t$, with
$H_{QDM}\left|\varphi\right>=0$. The dimer configuration corresponding
to $\left|\varphi\right>$\ turns out to constitute a topological
sector of its own. (Two configurations belong to the same topological
sector if one can be obtained from another by strictly local
rearrangements of the dimers.\cite{Rokhsar88})

As one decreases $v$\ through $t$, the ground state moves into another
sector, which contains an exponentially large number of dimer
configurations. The two candidate phases in this sector are depicted
in Fig.~\ref{fig:phases}b and c; these are the plaquette and columnar
valence-bond solids, respectively. In fact, for $v/t\rightarrow -
\infty$, one can see that the ground state will be the columnar state,
as this maximises the number of flippable plaquettes favoured by the
potential term. 

The point $v/t=1$ is the RK point where each equal amplitude 
superposition over a winding number sector is a ground state. An
analysis in terms of height representations\cite{henleyun} shows 
that there is a diverging correlation length as
one approaches this point from $v < t$ and that the critical theory
is Gaussian. In the same language the two candidate states mentioned
above for $v < t$ are flat but the competition between them cannot
be settled in the same analysis. We now turn to an alternative
mapping of the physics of the QDM which will allow us to settle
that question by computation.

\section{Useful mappings and numerical method}
\label{sect:maps}

This alternative, duality, mapping 
crisply distinguishes between the different phases.  
This mapping
takes the QDM in $d=2$ onto a classical, stacked, frustrated,
anisotropic Ising magnet in $d=2+1$ on its {\em dual}
lattice.\cite{mcs2000} The Hamiltonian for that model reads: \bea
\beta_C H_{Ising}= K^\xi\sum_{\left<ij\right>}\sigma_i\sigma_j-
K^\tau\sum_{\left<ii^\prime\right>}\sigma_i\sigma_{i^\prime}+
\beta_Cv_C \sum_i\delta_{B_i,0} \nonumber \ .
\eea

Here, the $\sigma$\ are Ising variable defined on the sites of a
stacked triangular lattice; the sum on $\left<ij\right>$\ runs over
nearest neighbour pairs in the plane, whereas the one on
$\left<ii^\prime\right>$\ is over pairs in adjacent layers. $B_i$\ is
the in-plane exchange field experienced by spin $i$; if it is zero,
the corresponding dimer plaquette is flippable.

To generate equivalent Hilbert spaces, one has to take the limit of
infinite exchange in the planes, $K^\xi\rightarrow+\infty$, as there
is a one-to-one correspondence between the hardcore dimer coverings on
the hexagonal lattice and the Ising model ground states on the
triangular lattice, up to a global spin reversal.\cite{fn-dual}

The equivalence then holds in the scaling limit
$K^\tau\rightarrow+\infty$, with the quantum inverse temperature
$\beta_Q$\ given by $\beta_Q t= \exp(2 K^\tau)/N\equiv\lambda/N$,
where $N$\ is the number of stacked layers, so that the zero
temperature limit corresponds to a system with infinite extent in the
stacking direction. The conversion of parameters between the classical
(C) and quantum (Q) problem proceeds via the formula
$v_Q/t=\beta_{C}v_C\lambda$.  In the following, the quoted values of
$v/t$\ are to be understood as referring to the quantum problem. Note
that $\lambda$\ (which we will quote in the following) quantifies the
discretisation error -- it gives a rough measure of a typical
correlation length in the stacking direction.

For the case $v=0$, this model has been studied in the past by several
groups. In an influential piece of work, Blankschtein and
coworkers\cite{Blankschtein84} have carried out a
Landau-Ginzburg-Wilson analysis for this model, which uncovered an
$XY$-symmetric action with an (in $d=2+1$\ dangerously irrelevant)
six-fold clock term, which breaks the $XY$\ symmetry at sixth
order. 

The ordering pattern obtained by this method is a three-sublattice
ordering pattern; depending on the sign of the six-fold clock term,
the three sublattices have the ordering pattern $(+M,0,-M)$, or
$(+m,-n,-n)$, where the amplitudes $M$, $m$\ and $n$\ are undetermined
in the most general scenario.\cite{fn-naive}
 Translating these back to dimer
correlations, one finds that the former pattern corresponds to the
plaquette VBS, whereas the latter corresponds to
the columnar VBS.

Simulating $H_{Ising}$\ is evidently straightforward in principle
using classical Monte Carlo simulations in $d=3$. The only
complication arises from the scaling limit which has to be taken,
requiring long correlation lengths in the stacking direction. Here, we
use a cluster algorithm,\cite{ferswen} the implementation of which
is easy in Ising language, but would have been rather hard to guess at
for simulations of a stacked dimer model.  The mapping onto the
stacked classical Ising model, together with the cluster algorithm, is
what enables us to simulate system sizes which are substantially in
excess of those treated so far in numerical studies of the QDM.

\section{Numerical results}
\label{sect:numres}

In this section, we display the results of our numerical simulations,
which we argue demonstrate the existence of a phase transition from
the columnar to the plaquette VBS around $v/t=-0.2\pm 0.05$. 

Since we know that the columnar phase is encountered in the limit
$v/t\rightarrow -\infty$, we can rephrase the question of whether
there also is the plaquette phase by asking whether there is a phase
transition as $v/t$\ is increased towards $+1$.  From the discussion
in the previous section, it is apparent that, in spin language, the
plaquette state has zero magnetisation whereas the columnar state does
not. All we therefore have to do is to look for restoration of the
Ising symmetry to discover whether there exists a plaquette
phase. This we do in the following section.

\begin{figure}
\epsfxsize=3.in
\centerline{\epsffile{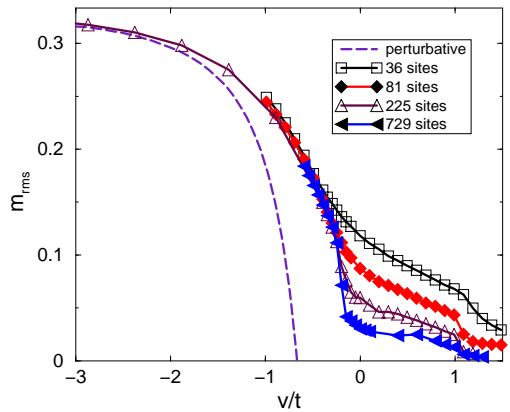}}
\caption{ $m_{rms}$\ as a function of $v/t$, in the sector containing
the columnar and plaquette phases. In this sector, the actual
transition at $v/t=1$\ to a state with strictly zero $m_{rms}$\ shows
up as a depression thereof; $\beta_Qt=0.083$, $\lambda=10$.}
\label{fig:magsqofV}
\end{figure}

To give an impression of the general phase diagram, consider
Fig.~\ref{fig:magsqofV}, where we plot the root-mean square
magnetisation, $m_{rms}$, of the equivalent Ising model, for small to
moderate system sizes over a broad range of the parameter $v/t$.  We
quote system sizes in terms of number of sites. This is the number of
unit cells of the hexagonal lattice, which equals the number of
dimers, and also the number of spins of the dual triangular
lattice. From this plot, a number of important features are already
visible.

Firstly, for large $-v/t$, $m_{rms}$\ approaches its limiting value
within the ground states of the Ising model, which is 1/3. Deviations
are well-captured perturbatively in $t/v$, as depicted by the dashed
line, which shows the lowest order result. Just to the left of
$v/t=0$, one can witness the vanishing of $m_{rms}$, which gets
sharper with increasing system size.

Finally, at the Rokhsar-Kivelson point, $v/t=1$, there is a kink in
$m_{rms}$, which indicates the further transition to the staggered
state. The staggered state has strictly zero magnetisation, but since
it is in a different sector from the other two states, this shows up
as a depression of $\mr$\ as $v$\ is swept through $t$,
as the system
tries to accomodate local staggered correlations within the wrong
topological sector as best it can.

\begin{figure}
\epsfxsize=3in
\centerline{\epsffile{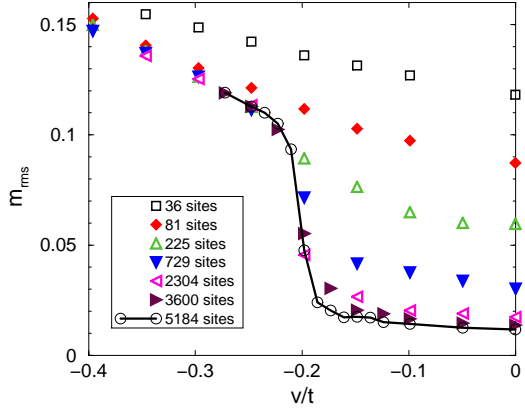}}
\caption{ Enlarged view of phase transition region. Note the first
order nature of the transition. The solid line is through the 
points
for to the largest system size. $\beta_Qt=0.083$, $\lambda=10$.}
\label{fig:magsqcrit}
\end{figure}

To get a handle on the details of the transition, we examine the same
plot for a much narrower range of $v/t$\ near the vanishing of $\mr$,
for a wider range of system sizes, as depicted in
Fig.~\ref{fig:magsqcrit}. As the system size increases, the transition
sharpens up into a discontinuous drop in $\mr$. This drop separates
the region on the left with $\mr$\ decreasing with an almost constant
slope, from that on the right, with $\mr$\ being near constant and
close to zero. The phase transition thus appears to be of first order,
as will be discussed in more detail below.

\begin{figure}
\epsfxsize=3in
\centerline{\epsffile{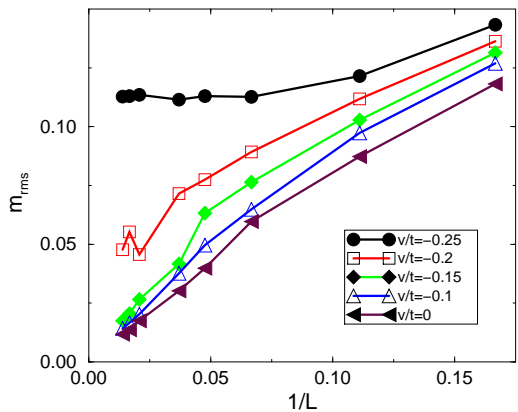}}
\caption{ Scaling of $\mr$\ as a function of $L^{-1}$, the inverse of
the linear system size. $\beta_Qt=0.083$, $\lambda=10$.}
\label{fig:msqscalesize}
\end{figure}

To underline this result, we plot the scaling of $\mr$\ as a function
of inverse linear system size for a number of values of $v/t$\ near
the transition in Fig.~\ref{fig:msqscalesize}. This plot shows that
$\mr$\ settles down to a nonzero value on the left of the transition,
at $v/t=-0.25$, whereas it scales to zero on the right, for
$v/t\geq-0.15$. The transition is located around $v/t=-0.2$, where the
scaling appears inconclusive. From this, we think it is conservative
to estimate the transition point between the two phases to be located
at $v/t=-0.2\pm0.05$.

We conclude this section by addressing potential systematic errors
arising from the introduction of the discretisation in the stacking
direction, since the mapping to the quantum dimer model is exact in
the continuum limit only.

\begin{figure}
\epsfxsize=3in
\centerline{\epsffile{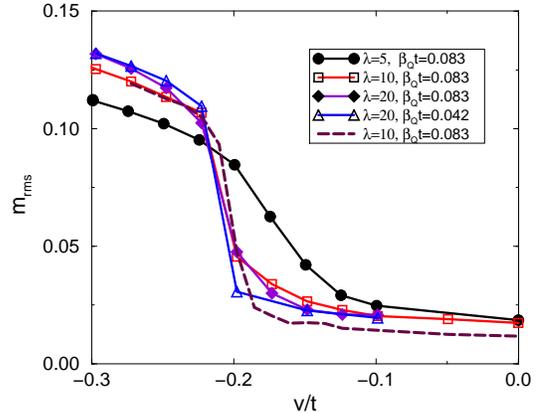}}
\caption{
Development of $\mr$\ as a function of $\lambda$\ and $\beta_Q$. 
The dashed line is for 5184 sites, the others are for 2304 sites.
Reducing the discretisation error (increasing $\lambda$), 
and lowering the quantum temperature (increasing $N$) sharpen 
up the transition.}
\label{fig:msqscaletemp}
\end{figure}

In Fig.~\ref{fig:msqscaletemp}, we show the plots of $\mr$\ vs.\
$v/t$\ for a system of 2304 sites using different couplings in the
stacking direction, $K^\tau$, thus varying $\lambda$, at a fixed
quantum temperature. It can be seen that the transition sharpens up as
$\lambda$\ is increased, but moves only little as $\lambda$\ changes
from 10 to 20. As the quantum temperature is lowered by a factor of
two at $\lambda=20$, the transition sharpens further but again does
not move significantly. These effects are therefore certainly within
the error bars we give for the value of the critcal $v/t$. The case of
the largest system we have studied (also displayed in
Fig.~\ref{fig:msqscaletemp}) clearly also falls into this range.

We note that the absence of finite-size effects at $v=0$, upon
increasing the number of layers, $N$, at fixed $\beta_C$\ and $L$,
implies the existence of a gap in this part of the phase diagram. This
is not surprising since at that point, we are far away from the phase
transition, which is first order at any rate. However, this
observation makes the existence of a gapless excitation at this point,
suggested in Ref.~\onlinecite{orland}, seem rather unlikely. More
generally, our results fit snugly into the expectations from 
the height representation analysis as well the analysis of the
transverse field Ising models (see below as well) and so there
seems little doubt that the analysis in Ref.~\onlinecite{orland}
is flawed.

\section{The phase diagram}
\label{sect:phasedia}

\begin{figure}
\epsfxsize=3in
\centerline{\epsffile{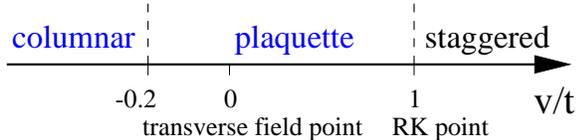}}
\caption{Phase diagram of the quantum dimer model on
the hexagonal lattice. The nature of the ordered phases is indicated 
above the axis. }
\label{fig:hexdimerphase}
\end{figure}

The phase diagram we have thus obtained is depicted in
Fig.~\ref{fig:hexdimerphase}.  The columnar-plaquette phase transition
is of first order, whereas the one at the RK point is a second order
one, albeit with the somewhat peculiar feature that, coming from the
right, it appears to be first order as
no fluctuations are visible leading up to the critical point.
However, coming from the left, a gap closes, giving rise to the
gapless resonon excitations.\cite{Rokhsar88}

There are a number of theoretical reasons which lead us to conclude
that the transition from plaquette to columnar VBS is first
order, as the simulations suggest. Within the framework of
the Landau-Ginzburg-Wilson theory,\cite{Blankschtein84} the critical
point corresponds to the vanishing of the coefficent of the six-fold
clock term, so that the system could in principle fluctuate between
all the degenerate $XY$\ states (including the columnar and plaquette
ones) without encountering any barriers. However, higher `harmonics'
(clock terms) will presumably come into play as they are unlikely to
vanish at exactly the same point as the leading one; it is these which
will prevent the barriers between the plaquette and columnar state
from vanishing.

Further, we note that the symmetry groups of the two VBSs are not such
that one of them is a subgroup of another, which would be a criterion
within Landau theory for a continuous transition. This is in fact a
somewhat subtle point as both phases break translational symmetry and
retain a six-fold rotational symmetry. However, when trying to form
domains of one phase within another, it turns out that the centres of
rotational symmetry lie in distinct places for the two phases.

This point, incidentally, is somewhat simpler in the square lattice,
where the columnar phase breaks translational symmetry in one
direction and also rotational symmetry, whereas the plaquette phase
breaks translational symmetry in both directions but retains a
four-fold rotational symmetry.

\section{Stacked magnets}
\label{sect:stacked}

Our simulations apply equivalently to the hexagonal dimer model and to
the stacked triangular magnets. We therefore briefly digress here to
note some implications of our results to the latter system, on which a
good deal of work has been done, in great part inspired by the
existence of experimental compounds realising this model; for a review
of both theory and experiment, see Ref.~\onlinecite{collinsrev}.

Recall
that $H_{Ising}$\ at $v_C=0$ reduces precisely to this model; there,
the presence of the plaquette VBS corresponds to the three-sublattice
$(+M,0,-M)$\ ordering pattern for the triangular magnet.  This agrees
with the results of Refs.~\onlinecite{mats87,heino89,kim90}, whereas
it partially disagrees with
Refs.~\onlinecite{Blankschtein84,Netz84}. 

This result is somewhat surprising, as even in the low-temperature
limit, it appears that one ends up with only a partially ordered state
(spins on one sublattice are still equally likely to point up or down
in this phase), rather than the apparently more fully ordered state
$(m,-n,-n)$. However, note that in either case, fluctuations are
present down to zero temperature -- in fact, these states are
stabilised by those fluctuations in the first
place.\cite{Coppersmith85,mcs2000} In the absence of fluctuations, the
energy of any ferromagnetically stacked ground state configuration of
the triangular magnet would be the same. Such a selection of an
ordered state by fluctuations is known as order by
disorder.\cite{Villain80} It has the feature that, although weak
fluctuations are needed for stabilising the state, their strengthening
will lead to a melting of the order they themselves established in the
first place.

In the present case, it now so happens that the intermediate phase
with a disordered sublattice can benefit from the fluctuations, and
survive them. As fluctuations are suppressed (e.g. by adding a
negative $v$, or a magnetic field, see below), one enters the phase
with a higher degree of ordering.

We can be reasonably confident that upon lowering the temperature even
further, there will not be a transition to the $(m,-n,-n)$, mainly
because the critical $v/t$\ seems to move very little, if at all in
the right direction, as the temperature is lowered. Nonetheless, we
are not entirely clear how to resolve the discrepancies with
Refs.~\onlinecite{Blankschtein84,Netz84}.  As for the hard-spin
mean-field theory,\cite{Netz84} it is conceivable that the
fluctuations are somewhat underestimated there, thus landing it on the
wrong side of the fine dividing line between the two states. At any
rate, we have explicitly simulated temperatures lower than the
expected transition temperature, and found no transition: for a system
of size $27\times27\times1024$\ spins at $K^\tau=2.3$\
($\lambda\simeq100$), we find sublattice root-mean square
magnetisations of $(0.95,0.12,0.95)$. The early simulations by
Blankschtein \etal\cite{Blankschtein84} may have run into problems as
ergodicity is lost for single-spin updating at low temperature with an
increasing correlation length in the stacking
direction,\cite{fn-paras} a problem only since remedied with the
development of more advanced Monte Carlo simulation
technology.\cite{ferswen,fn-cluster} However, our results are fully
consistent with the mean-field analysis of
Ref.~\onlinecite{Blankschtein84}, provided the coefficient of the
clock term remains of the same sign throughout the ordered phase.

\begin{figure}
\epsfxsize=2.9in
\centerline{\epsffile{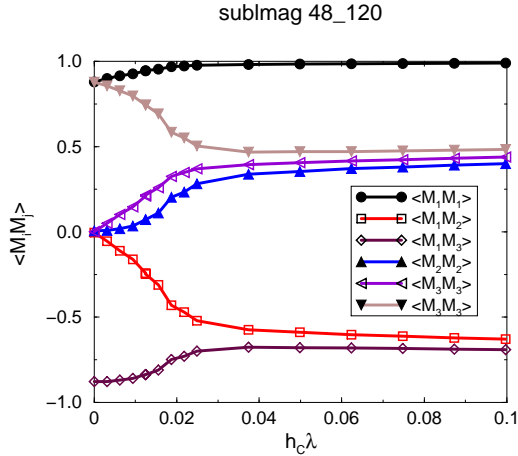}}
\caption{Correlation matrix of sublattice magnetisations
$(M_1,M_2,M_3)$, with $M_1>M_2>M_3$, as a function of magnetic field,
for a stack of 120 layers of 2304 sites. $K^\tau=1.15$, so that
$\lambda=10$. For $h=0$\ one finds the $(+M,0,-M)$\ phase, which
rapidly gives way to the $(m,-n,-n)$\ phase.}
\label{fig:sublmag}
\end{figure}

To illustrate the closeness of the two phases,\cite{Netz84,plumer95} 
in
Fig.~\ref{fig:sublmag} we display the sublattice correlation matrix
for a system of 2304 spins in a stack of height 120.  It can be seen
that upon application of a small magnetic field, one leaves the
$(+M,0,-M)$\ phase and enters the $(m,-n,-n)$\ one. Note that in the
classical model, the proximity of the phases is artificially enhanced
as the strength of the field is effectively multiplied by $\lambda$\
as the correlation length in the stacking direction increases. For
this reason, the abscissa of the plot is scaled up by $\lambda$.

\section{Relation to the Heisenberg antiferromagnet}
\label{sect:hexag}

The QDM was conceived to describe the physics of a
$S=1/2$\ Heisenberg antiferromagnet in a phase dominated by nearest
neighbour singlet bonds (valence bonds, or dimers). The question thus
arises under what circumstances a VB phase is energetically
competitive compared to the Neel state.  Note that a spin can form a
valence bond with only one of its neighbours, whereas it gains energy
from all its antialigned neighbours in a Neel state. It is via the
resonance moves, captured by $\hat{T}$, that it might make up the
energy difference.

One step towards destabilising the the Neel state is to maximise
quantum fluctuations, that is to say, consider spin $S=1/2$\ systems.
The original idea of Anderson\cite{Fazekas74} was to choose a
frustrated (triangular) magnet, as this does not permit a Neel state
in the first place.  In addition, in a lattice with low coordination
the disadvantage of each spin forming only one VB is relatively less
severe, thus favouring a VB state.

Moreover, it is advantageous not to have closed loops which are very
short. This can be seen from within the framework of the
QDM,\cite{Rokhsar88} which, formally, is an expansion in the
overlap between distinct VB configurations. This overlap is
exponentially small in the number of VBs in which they differ. The
shortest closed loop of even length on the lattice on which the VBs
reside determines the lowest order term in this expansion. This length
is four for the square and triangular lattice, but six for the
hexagonal lattice, thus favouring the latter. 
Furthermore, introducing
frustrating further-nearest neighbour exchanges yields a smaller
prefactor (but not a smaller expansion parameter) in the RK expansion.

The hexagonal lattice with frustrating further neighbour
exchanges\cite{matt94} would thus seem to be a good candidate for
being described by some sort of QDM. In fact, this expectation appears
to be borne out by very recent work of J. B. Fouet \etal,\cite{fouet}
who did exact diagonalisation studies on a 
$J_1$--$J_2$--$J_3$\ $S=1/2$\
Heisenberg model on the hexagonal lattice. They indeed found valence
bond phases. For frustrating $J_2/J_1=0.4$, they found a staggered
phase, which gives way to an (at least short-range ordered) columnar
or plaquette phase around $J_2/J_1=0.3$; the numerics, while not being 
entirely
conclusive, is suggestive of the latter phase.\cite{fouetpriv} 

This is in keeping with the QDM analysis, which indeed suggests that
this transition corresponds to crossing the RK point between the
plaquette and staggered phases.  We should note, however, that
carrying out the perturbation theory within the dimer manifold in the
spirit of Ref.~\onlinecite{Rokhsar88} does not place the point $v=t$\
in between those two values of $J_2/J_1$, and so a more microscopic
prediction of the properties of Heisenberg models will probably
require going beyond this by including any renormalizations needed to
obtain an effective QDM with nearest neighbor bonds only.

\section{Conclusions}
\label{sect:concl}

We have mapped out the phase diagram of the QDM on the hexagonal
lattice. We have established the existence of a plaquette VBS
intermediate between a columnar and a staggered one.  This was
achieved by combining a duality mapping to a classical model with a
Monte Carlo cluster algorithm. In the process, we were also able to go
well beyond the limitations of previous numerical work on the stacked
triangular magnet to confirm the nature of the low temperature phase
in that model.  Based on the close connections between the properties
of dimer models on the hexagonal and square lattices,\cite{mcs2000} we
expect the square lattice QDM, which can be obtained from similar
simulations on a stacked magnet,\cite{sachjala} to behave in an
analogous manner. As detailed in Sect.~\ref{sect:hexag}
for the hexagonal lattice, this could
shed some light on the properties of a $J_1-J_2$\ square lattice
Heisenberg model,\cite{kotovj12} for which the derivation of the RK
model is totally analogous. Nevertheless the competition between the
plaquette and columnar phases is a matter of microscopic detail and so
a thorough study of the latter would appear to be in order
definitively to settle the issue.  We expect to tackle this question
in future work.

\section*{Acknowledgements}
We would like to acknowledge conversations with J.B. Fouet, C. Henley,
D. Huse, C. Lhuillier, S. Sachdev and P. Sindzingre. The work was
supported in part by grants from the Deutsche Forschungsgemeinschaft,
NSF grant No. DMR-9978074, the A. P. Sloan Foundation and the David
and Lucile Packard Foundation.


\begin{references}
\bibitem{fn-naive} In fact, the naive LGW treatment gives $n=m/2$\ and
hence both patterns with zero magnetisation. However, for 
the $(+M,0,-M)$\ but not for the $(+m,-n,-n)$\
pattern, this property is protected by the sublattice symmetry,
so that generically $n\neq m/2$.
We have explicitly checked in our numerics that the Ising symmetry
breaking transition is indeed the one between these
two patterns.
\bibitem{fn-paras} Note, also, that the microscopic parameters used in
the simulations are not exactly the same. In our simulation, we took
the limit of infinite spatial coupling.
\end{references}
\end{document}